\documentclass[aps,pra,twocolumn,10pt,longbibliography,nobibnotes,floatfix]{revtex4-2}

\usepackage[T1]{fontenc}
\usepackage[utf8]{inputenc}
\usepackage{lmodern}
\usepackage{amsmath,amssymb,amsfonts,bm,dsfont}
\usepackage{ulem}
\usepackage{graphicx}
\usepackage{xcolor}
\usepackage{hyperref}
\hypersetup{
colorlinks=true,
linkcolor=blue!50!black,
citecolor=blue!50!black,
urlcolor=blue!50!black
}

\newcommand{\bra}[1]{\langle #1|}
\newcommand{\ket}[1]{|#1 \rangle}
\newcommand{\ketbra}[2]{|#1\rangle\!\langle#2|}

\newcommand{\av}[1]{\langle #1 \rangle}
\newcommand{\modsq}[1]{| #1 |^2}

\newcommand{\Enq}{\mathcal E_{N}^{(q)}}
\newcommand{\En}{\mathcal E_N}
\newcommand{\Q}{\mathcal Q}

\begin{document}

\title{Scalable quantum resources with short-range interacting spin-1/2 chains}

\author{Marcin P{\l}odzie\'n}
\affiliation{Qilimanjaro Quantum Tech, Carrer de Vene\c{c}uela 74, 08019 Barcelona, Spain}

\author{Jan Chwede\'nczuk}
\affiliation{Faculty of Physics, University of Warsaw, ulica Pasteura 5, 02-093 Warszawa, Poland}

\begin{abstract}
  The dynamical generation of quantum resources, such as many-body entanglement, Bell correlations or spin squeezing, can be achieved via one-axis twisting (OAT) dynamics, which require all-to-all couplings.
  However, current digital and analog quantum simulation platforms natively provide short-range or power-law couplings that decay too quickly for this purpose.
  We demonstrate that two spin-$\tfrac12$ chain models---a staggered nearest-neighbor XXX chain and a long-range XXZ chain---develop 
  an effective OAT nonlinearity when projected onto the symmetric sector.
  We show that these dynamics generate metrologically useful 
  spin-squeezed states and Greenberger-Horne-Zeilinger coherences that ensure violation of many-body Bell inequalities.
  We confirm the accuracy of this mapping by comparing it to the exact dynamics and demonstrate that the generated correlations can be read out using a single probe qubit.
  The resulting dynamics can be simulated with analog and digital quantum simulators. 
\end{abstract}

\maketitle

\section{Introduction}Programmable quantum simulators now offer precise control over large, interacting spin systems and high-fidelity operations. 
However, generating strongly entangled or Bell-correlated states remains challenging since these platforms natively realize short-range or power-law interactions rather than collective, all-to-all couplings. 
In this context, the one-axis twisting (OAT) model~\cite{KitagawaUeda1993} serves as a reference. 
It is used to generate a hierarchy of entangled states. Early on,  it yields metrologically useful spin-squeezed states which enable 
quantum-enhanced precision in interferometric sensing~\cite{Wineland1992,Wineland1994,KitagawaUeda1993}. 
Later, oversqueezed non-Gaussian states are created, followed by the Greenberger-Horne-Zeilinger (GHZ) states at the optimal time~\cite{KitagawaUeda1993,Pezze2018RMP,Plodzien2022}.

We pose the following question: Are short-range interactions characteristic of modern architectures a fundamental shortcoming that prevents the generation of macroscopic entanglement and nonlocal correlations?
We demonstrate that effective long-range interactions, which are accessible through native spin-chain dynamics, 
provide access to the entire hierarchy when the Schrieffer–Wolff transformation (SWT) maps the system onto the symmetric Dicke manifold.
We identify the conditions under which a finite magnon gap isolates the zero-momentum collective mode. Virtual excitations across the gap yield 
effective Lipkin-Meshkov-Glick (LMG) interactions~\cite{Lipkin1965I}, equivalent to those characteristic of OAT dynamics.

More quantitatively, we compare the resources and correlations  generated by short- or power-law interactions, to the outcome of an indeal OAT dynamics, in particular of 
the GHZ state~\cite{Cavalcanti2019PRA,CavalcantiPRL2007,CavalcantiPRA2011,HePRA2011,Niezgoda2020,Niezgoda2021,Chwedeczuk2022,Plodzien2022,Plodzien2024PRA,Plodzien2024PRR,
  Plodzien2025ROPP,Plodzien2025PRA,HernandezYanes2025}. 
The latter is characterized by a maximal quantum coherence between two ``macroscopically'' different components, which we denote as ``GHZ coherence''. 
It is a source of
correlations that are incompatible with any local realistic model~\cite{Bell1964,CHSH1969,Mermin1990,Brunner2014RMP}.  
 
We apply this technique to some microscopic models which can be realized both in gate-based (digital) quantum processors and analog quantum simulators. We start with the 
 staggered XXX chain with nearest-neighbour interactions and { we extend the analysis to the} power-law XXZ model. 
 { This illustrates the method's universality and enables us to connect with modern qubit platforms characterised by short-range interactions.}
In both cases, we demonstrate that effective long-range nonlinearity generates many-body Bell correlations.
Over longer time, the system violates a many-body Bell inequality~\cite{Tura2014Science,Tura2015AnnPhys}.
Moreover, we show that metrologically useful spin-squeezed states can be generated with nearest-neighbour interactions on quantum circuits without a need for gate optimization, contrary to recent proposals based on variational quantum circuits~\cite{Kaubruegger2019VSS,Sun2023Variational1D,Lyu2023VariationalLR,Song2025DipoleSqueezing}.
We discuss the regimes of validity of the SWT 
by demonstrating when the exact lattice dynamics---continuous-time analog evolution or Trotterized digital circuits---closely match the effective collective description.  
Furthermore, we demonstrate that the resulting correlations can be experimentally verified by collectively coupling the spin chain to a single probe qubit~\cite{PlodzienChwedenczuk2025_probe_qubit}.
{ This method enables the correlations of any depth of entanglement or non-locality to be read out, eliminating the need for cumbersome measurements on a multi-spin system.}

{Note that the SWT has been discussed in the context of the transition from an effective short- to long-range interaction in Ref.~\cite{Gietka2020}. 
  Here, we present a step-by-step derivation of this important result and discuss the range of its validity.
  This concept is both timely and relevant to the quantum technologies community. The effective generation of spin-squeezed states in spin lattices of various topologies and dimensionality 
  has been discussed in~\cite{PhysRevA.109.023326}. 
  More recently, the experimental realisation of ensembles of Rydberg-dressed atoms has demonstrated the importance of methods based on virtual long-range interaction~\cite{PhysRevLett.131.063401}
  Finally, the methods discussed in this paper can be used to derive an effective two-axis counter-twisting Hamiltonian~\cite{koyluoglu2025}
  
  This manuscript is organised as follows. First, in Section~\ref{sec.intro}, we derive the effective Hamiltonians for some prominent examples of spin chains. 
  Next, in Section~\ref{sec.res}, 
  we demonstrate how the SWT generates entanglement and Bell correlations in these systems. We also demonstrate how these correlations can be probed using a single qubit connected to the chain.
  Further discussion and conclusions are presented in Section~\ref{sec.disc}. See the Appendices for the detailed, step-by-step derivation of all the relevant results.

}

\section{Effective Hamiltonians}\label{sec.intro}
We begin with the general Hamiltonian of an $N$-particle spin-$\tfrac{1}{2}$ chain
\begin{align}\label{eq:LR_generic}
  \hat{H}=-\frac{1}{2}\sum_{i\neq j=1}^NJ(r_{ij})\hat{\vec S}_i \hat{\vec S}_{j}+\sum_{j=1}^N\sum_{\alpha} h_j^{\alpha}\,\hat S_j^\alpha,
\end{align}
where the first sum runs over all pairs $i\neq j$ and $\alpha=x,y,z$.
The two-body coupling  is  distance-dependent, i.e., $r_{ij}=|i-j|$, while 
the operator for the $i$-th spin is $\hat S_i^\alpha$. 
To perform the SWT, we first introduce the discrete Fourier transform of a quantity $f$, $\tilde f_\alpha(q)=\frac1{\sqrt N}\sum_{j=1}^{N}f_j^\alpha e^{iqj}$. 
The switching to the momentum space is important because $q$ distinguishes the low-energy ($q=0$) fully symmetric part of the spectrum of $\hat H_0$  from
the single-magnon
excitations characterized by the energy $\varepsilon(q)=\tfrac{1}{2}[\tilde J (0)-\tilde J(q)]$, where $\tilde J(q)=\sum_{r=1}^{N-1}J(r)\,\mathrm{e}^{iqr}$.
The magnon gap is $\Delta=\min_{q\neq 0}\varepsilon(q)$.
The SWT can be applied if the couplings $h_j^{\alpha}$ are small, since the virtual excitations above the gap are driven by the operator $V$. This gives the effective Hamiltonian
\begin{align}\label{eq:H_LMG}
  \hat{H}_{\rm eff}=\sum_{\alpha,\beta}\chi_{\alpha\beta}\,\hat{S}_{\alpha}\hat{S}_{\beta},
\end{align}
where the collective spin operators are  $\hat{S}_{\alpha}=\sum_i\hat S_i^\alpha$. Only the symmetric part of the coupling tensor contributes (see the Appendix), i.e., 
\begin{align}
  \chi_{\alpha\beta}=\sum_{q}\frac{\tilde{h}_{\alpha}(q)\,\tilde{h}_{\beta}(-q)}{N(N-1)\varepsilon(q)}.
\end{align}
This result can be applied directly to some of the most notable subclasses of the Hamiltonian~\eqref{eq:LR_generic}.

First, we focus on the staggered XXX model
\begin{align}\label{eq:H_staggered_XXX}
  \hat H= -J_0\sum_{i=1}^{N}\hat{\vec S}_i \hat{\vec S}_{i+1} - h_z\sum_{i=1}^{N}(-1)^i \hat S_i^z,
\end{align}
where we assume periodic boundary conditions throughout.
The one-magnon dispersion relation of $\hat H_0$ is
$\varepsilon(q)=J_0(1-\cos q)$, and the corresponding energy gap equals $\Delta=\varepsilon(\pi)=2J_0$.
The Fourier transform of the staggered field has single component $\tilde h_z(q)=-h_z\sqrt{N}\,\delta_{q,\pi}$, and the effective collective model, Eq.~\eqref{eq:H_LMG}, reduces to the 
OAT Hamiltonian~\cite{Gietka2020, HernandezYanes2022, Hanes2023PRA}
\begin{align}\label{eq:H_OAT_effective}
  \hat H_{\rm eff}=\chi\hat S_z^2, \quad \chi=\frac{h_z^2}{2J_0(N-1)}.
\end{align}

Another example is the paradigmatic long-range anisotropic Heisenberg (XXZ) spin-$1/2$ model,
\begin{align}\label{eq:H_XXZ_main}
  \hat{H}=-\frac{1}{2}\sum_{i\neq j}J(r_{ij})\left(\hat{\vec S}_i \hat{\vec S}_j+\delta\hat{S}_i^z \hat{S}_j^z\right),
\end{align}
where $J(r_{ij})=J_0|i-j|^{-\gamma}$ sets the microscopic coupling scale ($J_0>0$),
$\gamma$ controls the range of power-law interactions,
and $\delta$ parametrizes the easy-axis anisotropy.
For $\gamma\leq 1$, the sum $\tilde{J}(0)=\sum_{r}J(r)$ grows extensively with $N$, so that the energy per spin diverges in the thermodynamic limit.

The  remedy to this problem is the Kac normalization~\cite{Kac1963}: rescale $J(r_{ij})\to J(r_{ij})/\mathcal{N}_{\gamma,N}$ with $\mathcal{N}_{\gamma,N}=\sum_{j\neq i}J(r_{ij})/J_0$. 
This gives $\tilde{J}(0)=J_0$ for every $\gamma$, so the effective nonlinearity in Eq.~\eqref{eq:H_OAT_XXZ} 
carries only the explicit Kac scaling $\chi\propto 1/(N-1)$; the interaction range enters only through the magnon gap that protects the symmetric manifold.
The model interpolates between the isotropic Heisenberg limit ($\delta=0$) and the strongly anisotropic Ising limit ($|\delta|\!\gg\!1$).

In contrast to the staggered XXX chain, where the OAT nonlinearity emerges as a second-order perturbation from the one-body field, the XXZ model has a symmetry-breaking anisotropy in the two-body coupling.
As a result, the effective collective Hamiltonian is obtained by applying a projection $\hat\Pi$
onto the symmetric Dicke manifold, for instance
\begin{align}\label{eq:Pi_SizSjz}
  \hat\Pi\hat S_i^z\hat S_j^z\hat\Pi=\frac{1}{N(N-1)}\left(\hat S_z^2-\frac{N}{4}\right)\hat\Pi,
\end{align}
which holds for any pair $i\neq j$ in the fully symmetric sector. The resulting Hamiltonian $\hat{H}_{\mathrm{eff}}=\hat\Pi\hat H\hat\Pi$, up to an additive constant, is
\begin{align}\label{eq:H_coll_XXZ}
  \hat{H}_{\mathrm{eff}}=-\frac{\sum_{i\neq j}J_{ij}}{2N(N-1)}\Big(\hat S^2+\delta\hat S_z^2\Big)\hat\Pi
\end{align}
Since $\hat S^2=S(S+1)\hat{\mathds1}$ in the symmetric subspace, we end up with the expression
\begin{align}\label{eq:H_OAT_XXZ}
  \hat H_{\mathrm{eff}}\simeq \chi\hat S_z^2,\qquad \chi=-\frac{\delta}{2}\frac{\tilde{J}(0)}{N-1}.
\end{align}
The collective description is adequate when the magnon gap pertubatively suppresses the leakage on the timescale $t\sim 1/|\chi|$ (see the Appendix).
The sign of $\chi$, controlled by $\delta$, determines the squeezing axis, though it does not affect the generation of Bell correlations.
The formula~\eqref{eq:H_OAT_XXZ} is exact within the symmetric manifold and holds for any value of $\delta$, but the full XXZ evolution drives the system out of this subspace when $\delta\neq 0$. 
In the two limiting regimes, the gap has distinct physical origins.

When the  anisotropy is small ($|\delta|\ll 1$), the dominant part of the Hamiltonian~\eqref{eq:H_XXZ_main} is SU(2)-invariant and strictly preserves the symmetry. 
Its one-magnon dispersion $\varepsilon(q)$ provides a finite gap that grows with the interaction range (decreasing $\gamma$), ensuring that the anisotropy 
$\parallel\!\!\hat{V}\!\!\parallel\propto\delta$ is small.
On the other hand, when the anisotropy is large ($|\delta|\gg 1$), 
the dominant Ising coupling $\hat{H}_0=-\frac{1+\delta}{2}\sum_{i\neq j}J_{ij}\hat{S}_i^z\hat{S}_j^z$ 
opens a gap $\Delta\sim(1+\delta)\tilde{J}(0)$  above the ferromagnetic ground states.
Second-order virtual spin-flip processes across this gap generate an effective OAT nonlinearity consistent with Eq.~\eqref{eq:H_OAT_XXZ}.
The complete derivation, including both limiting regimes and the XXZ dispersion relation, is presented in the Appendix.

\begin{figure}[t]
  \centering
  \includegraphics[width=0.9\columnwidth]{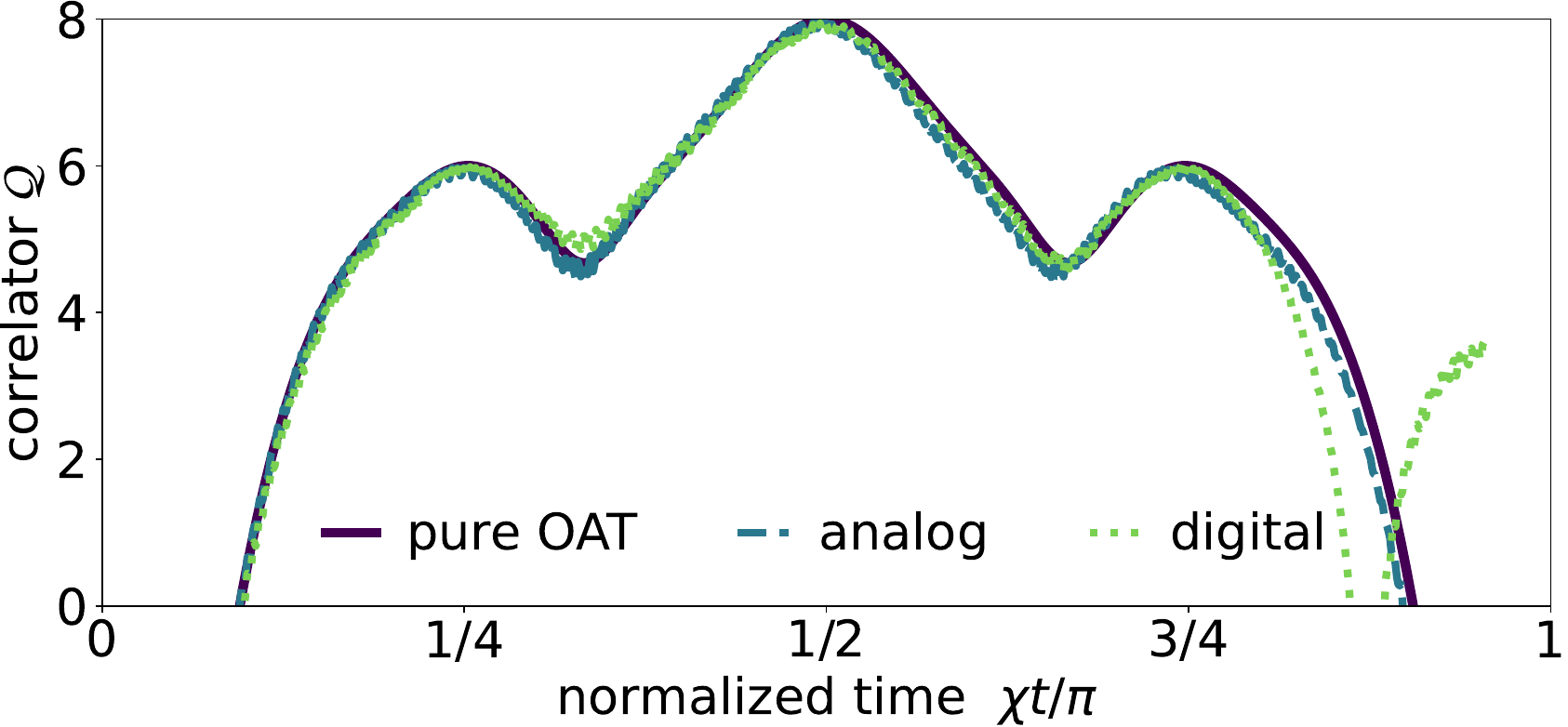}
  \caption{Time evolution of the Bell correlator $\Q(t)$ for $N=10$ spins and $h_z=0.15$ in the staggered XXX  chain.
    The solid line corresponds to the collective OAT results. The dashed line corresponds to the exact analog evolution under the XXX Hamiltonian obtained from full diagonalization. 
    The dotted line shows the Trotterized dynamics generated by the digital quantum circuit. The three curves coincide when time is rescaled by the effective OAT coupling.
    For weak staggered fields, all implementations agree and display a pronounced peak in $\Q(t)$, signaling the creation of GHZ-like correlations via the effective $\hat S_z^2$ twisting.}
  \label{fig:staggered_XXX_Q_vs_time}
\end{figure}

\section{Results}\label{sec.res}
We will now test the accuracy of the SWT. For each case, the initial condition is a fully polarized product of the +1/2 eigenstates of the $\hat S_x^{(i)}$ operators,
\begin{align}
  \ket{\psi(0)}=\ket{\uparrow}_x^{\otimes N}.
\end{align}
This state will be propagated using either an exact or an approximate symmetric Hamiltonian to yield  $\ket{\psi(t)}$.
To perform a quantitative comparison, we calculate the strength of the GHZ coherence, denoted as $\Enq$, for each case. 
It is the modulus squared of the most off-diagonal term of the density matrix $\ketbra{\psi(t)}{\psi(t)}$. This quantity is a witness of Bell correlations in quantum $N$-qubit systems, since
$\Enq\leqslant2^{-N}$ is an $N$-body Bell inequality [see Refs.~\cite{CavalcantiPRL2007,CavalcantiPRA2011,HePRA2011} and the Appendix] or equivalently $\Q=N+\log_2\Enq\leqslant0$. Hence $\Q>0$ signals the presence
of many-body Bell correlations. It reaches a maximal value of $\Q_{\rm max}=N-2$ for the GHZ state.

Furhtermore, we estimate the metrological utility of the above systems by imprinting a phase $\theta$ on the output state, $\ket{\psi_\theta(t)}=e^{-i\theta\hat S_y}\ket{\psi(t)}$. 
Next, we calculate the population imbalance probability $p_m(\theta)$, where $m=\frac12(n_\uparrow-n_\downarrow)$ is the difference between the number of spins in the up and down states.
This probability can be obtained by measuring each spin's configuration. 
However, as our recent findings indicate~\cite{gmpy-2sgh}, a single external probe qubit interacting with the $N$-spin system
can harness the full information about $p_m(\theta)$.

More specifically, 
before we ``connect'' the chain to the probe, we imprint the relative phase $\theta$ between the components of $\ket{\psi(t)}$  expressed in the basis of the symmetric Dicke states.
Specifically, we implement a series of three rotations
\begin{align}
  \ket{\psi_\theta(t)}=e^{-i\tfrac{\pi}{2}\hat{S}_x}e^{-i\theta\hat{S}_z}e^{-i\tfrac{\pi}{2}\hat{S}_y}\ket{\psi(t)}.
\end{align}
The first (time-wise) transforms the GHZ-like coherences that at the output of the dynamical protocols are aligned along the $y$-axis so that they lie along $z$ and are most
susceptible to the imprint of the phase $\theta$ (second rotation). The final rotation mixes the modes to produce the $\theta$-dependent signal.
Afterwards, the probe qubit is initialized in
\begin{align}
  \ket{\uparrow}^{(p)}_x=\frac1{\sqrt2}\left(\ket{\uparrow}^{(p)}_z+\ket{\downarrow}^{(p)}_z\right), 
\end{align}
where the index $(p)$ denotes the probe's degrees of freedom. The qubit is coupled to the chain through
\begin{align}
  \hat{H}_{\mathrm{int}}=\kappa\hat{S}_z\hat{S}_z^{(p)}.
\end{align}
and the composite (chain+probe) state evolves as
\begin{align}\label{eq:add_rot}
  \ket{\Psi(t,\theta)}=e^{-i\hat{H}_{\mathrm{int}}t}\bigl(\ket{\psi(\theta)}\otimes\ket{\uparrow}^{(p)}_x\bigr).
\end{align}
The quantity of interest is the off-diagonal terms of the probe's reduced density matrix
\begin{align}
  a(t,\theta)=\bra\downarrow\hat{\varrho}_p(t,\theta)\ket\uparrow,
\end{align}
where for brevity we used $\ket{\uparrow/\downarrow}\equiv\ket{\uparrow/\downarrow}^{(p)}_z$.
Since $\hat{H}_{\mathrm{int}}$ does not couple different eigenvalues of $\hat{S}_z$, the probe coherence is a Fourier series in the total magnetization $m$ of the chain,
\begin{align}
  a(t,\theta)=\frac{1}{2}\sum_{m=-N/2}^{N/2}p_m(\theta)\,e^{-i m\tau},
\end{align}
where $\tau=\kappa t$, $J=N/2$ is the total spin quantum number, and $p_m(\theta)$ is the probability that the chain in state $\ket{\psi(\theta)}$ has total magnetization $\hat{S}_z\ket m=m\ket m$. 
Sampling $a(t,\theta)$ at a discrete grid of times
\begin{align}
  \tau_k=\frac{2\pi k}{N+1},  \qquad k=0,1,\dots,N,
\end{align}
puts this relation in discrete Fourier-transform form. Inverting it yields
\begin{align}\label{eq:prob_pm}
  p_m(\theta)=\frac{2}{N+1}\sum_{k=0}^{N}a(\tau_k,\theta)e^{i\,\tfrac{2\pi k}{N+1}(m+\tfrac{N}{2})}.
\end{align}
To summarize, the procedure presented above gives access to the full set $\{p_m(\theta)\}$ using single-probe off-diagonal element $a$. For a detailed discussion, see~\cite{gmpy-2sgh}.

\begin{figure}[t!]
\centering
\includegraphics[width=\columnwidth]{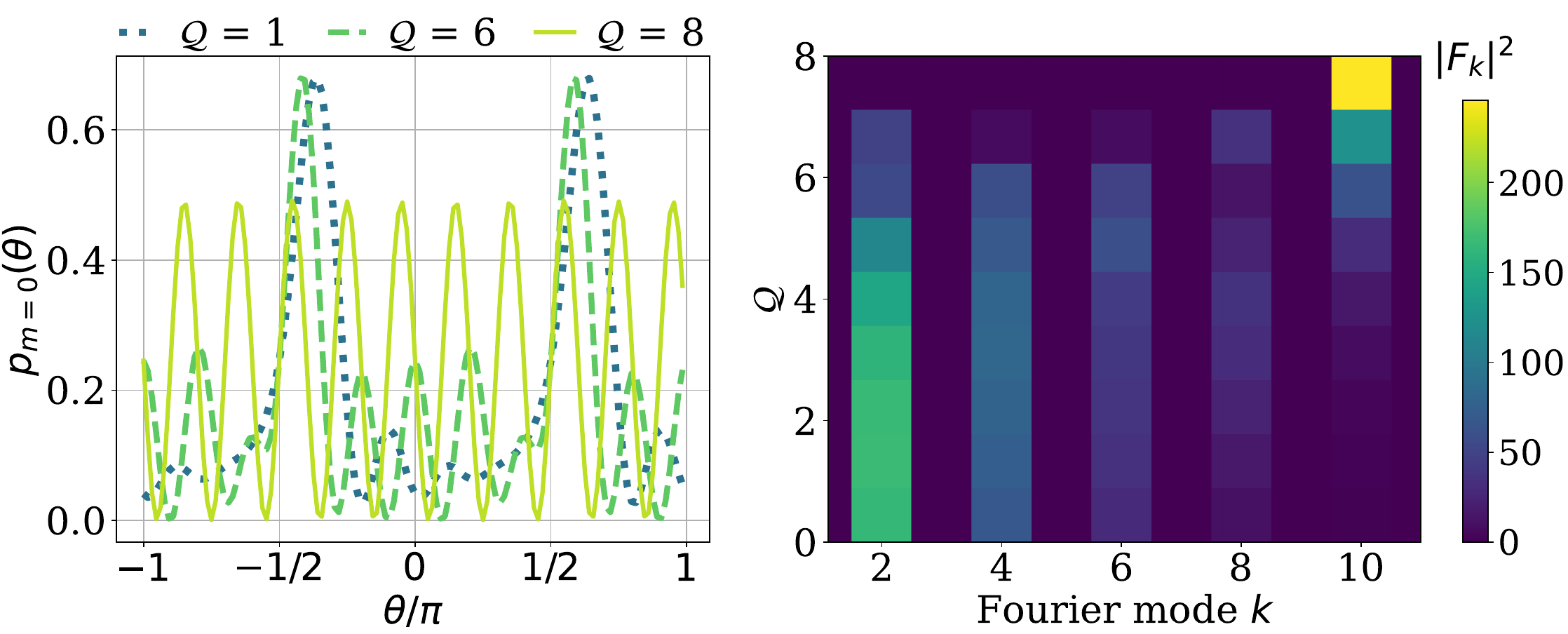}
\caption{Probe-qubit certification of the many-body Bell correlator for the staggered XXX model, Eq.~\eqref{eq:H_staggered_XXX}, with $N = 10$, 
  $h_z = 0.1$, and periodic boundary conditions.
  Left panel: probability $p_{m=0}(\theta)$ as a function of the imprinted phase $\theta$, shown for states with Bell correlators $\Q = 1, 6, 8$.
  Right panel: squared Fourier amplitudes $|F_k[p_0]|^2$ as a function of the harmonic index $k$. 
  In agreement with the parity constraint of the Wigner $d$-matrix, only even harmonics contribute. The GHZ-sensitive mode at $k=N$ grows systematically with $\Q$.}
\label{fig:Q_vs_k}
\end{figure}
 
The functions $p_m(\theta)$ provide information about the Dicke-basis matrix elements of the original state. Writing
\begin{align}
  \ketbra{\psi(t)}{\psi(t)}=\sum_{m,m'=-J}^{J}\varrho_{mm'}\ketbra m{m'},
\end{align}
we obtain that the probability from Eq.~\eqref{eq:prob_pm} is equal to
\begin{align}\label{eq:disc_four}
  p_m(\theta)=\sum_{m',m''}d_{m m'}d_{m m''}^*\varrho_{m'm''}\,e^{-i(m''-m')\theta},
\end{align}
where $d_{nm}=d^J_{nm}(\pi/2)$ are Wigner $d$-matrix elements that stem from the additional rotation through the $x$ axis present in Eq.~\eqref{eq:add_rot}.

Note that Eq.~\eqref{eq:disc_four} is, in fact, a discrete Fourier transform with coefficients given by the density matrix elements. 
By inverting this relation, i.e., calculating $\mathcal F^{(-1)}[p_m(\theta)]$ for some $m$ (for instance $m=0$ that for symmetric states gives the strongest signal),
and focusing on the term corresponding to the highest frequency (when $m=-m'=N/2$), we obtain
\begin{align}\label{eq:disc2}
  \varrho_{\frac N2,-\frac N2}=\frac{\mathcal F^{-1}[p_0(\theta)]}{N_{\theta}d_{0,\frac N2}^2},
\end{align}
where $N_\theta$ is the number of points on which $\theta$ is sampled. This way, a measurement on the single-qubit probe and a subsequent 
two-fold discrete Fourier transform, see Eq.~\eqref{eq:prob_pm} and~\eqref{eq:disc2} gives access to the GHZ coherence, and thus provides information
about the strength of Bell correlations in the spin chain. 
In particular, as entanglement strengthens during the OAT dynamics, the GHZ components begin to dominate the state $\ket{\psi(t)}$. 
This growing presence can be deduced from a Fourier transform of $p_m(\theta)$, which yields the magnitude of the terms oscillating at the highest frequency: $N\theta$.

First, we consider the staggered nearest-neighbor XXX model. One way to implement this type of dynamics is through continuous-time analog evolution, which is generated by the Hamiltonian 
from Eq.~\eqref{eq:H_staggered_XXX}.
Another approach involves a digital Trotter circuit. First, split $\hat H_{\mathrm{XXX}}$ into even and odd nearest-neighbor layers, $\hat H_{e/o}$, as well as a staggered-field layer, $\hat V$. 
Then, apply the second-order Suzuki--Trotter decomposition~\cite{Trotter1959,Suzuki1976,Lloyd1996} $\hat U(\delta t)=\hat U_V\left[\hat U_e\hat U_o\right]^2\hat U_V$,
where $\hat U_{e/o/V}$ are the corresponding evolution operators and $\delta t=t/n$ is the Trotter step. The exact dynamics is recovered when $n\to\infty$.

In Fig.~\ref{fig:staggered_XXX_Q_vs_time} we plot $\Q(t)$ for $N=10$:  the red-dashed 
line is the result of the full analog evolution~\eqref{eq:H_staggered_XXX}, the black solid curve is obtained with a bare OAT Hamiltonian~\eqref{eq:H_OAT_effective}, 
while the blue-dotted line is obtained with the Trotter circuit. All three curves display a pronounced peak $\Q\approx \Q_{\rm max}$, 
showing that the staggered XXX chain generates the GHZ correlations through the emergent OAT nonlinearity despite its short-range couplings.
The left panel of Fig.~\ref{fig:Q_vs_k} displays the central magnetization sector $p_0(\theta)$ for states with different $\Q(t)$. Clearly, as $\Q$ grows, the oscillations become faster. 
The right panel shows the squared Fourier amplitudes $\left|F_k[p_0]\right|^2$ as a function of the mode index $k$ and $\Q$.
In agreement with the parity constraint imposed by the Wigner $d$-matrix, only even harmonics appear. The GHZ-sensitive mode at $k=N$ increases monotonically with $\Q$, 
whereas all other remain comparatively small. This behavior reflects the growth of extremal Dicke (i.e., the GHZ) coherence, demonstrating that the probe-based protocol is a sensitive, 
experimentally feasible method of reading out Bell correlations.

\begin{figure}[t!]
  \centering
  \includegraphics[width=\linewidth]{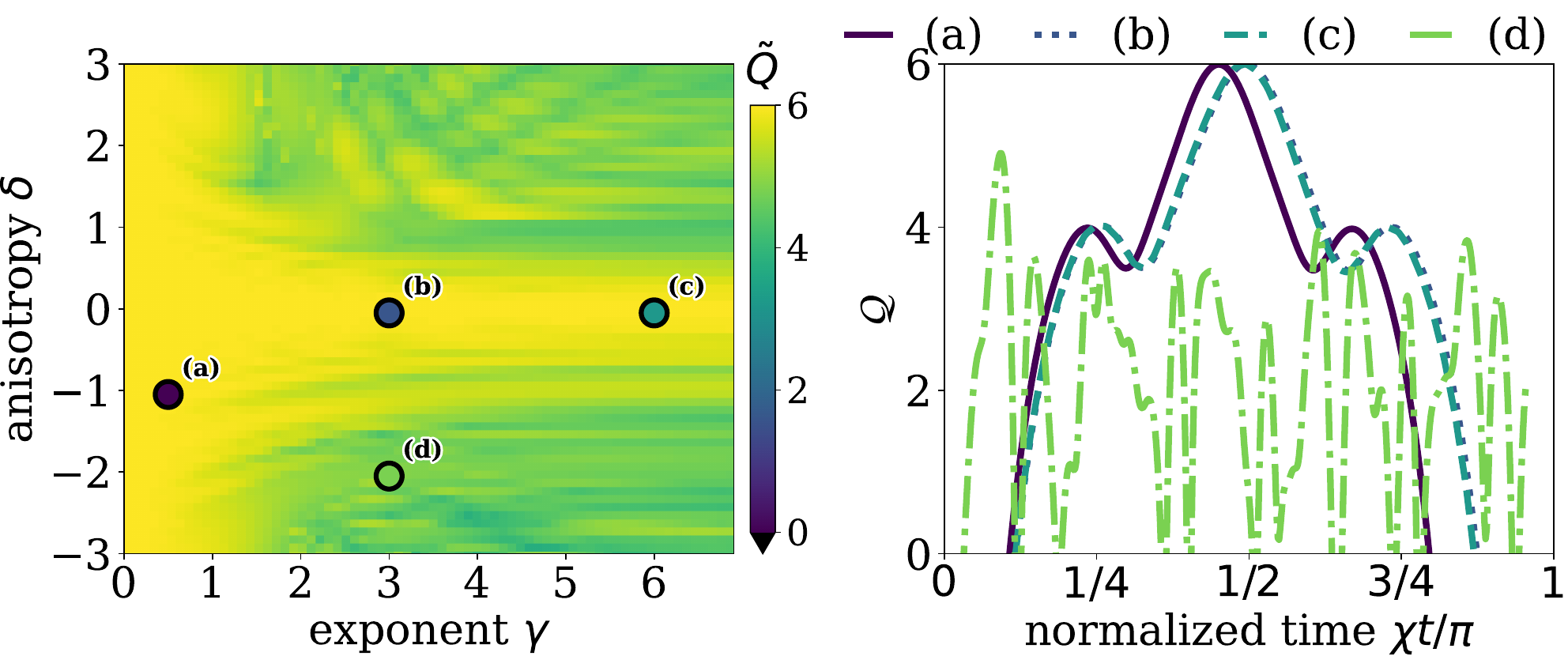}
  \caption{Bell correlations in the long-range XXZ model $\hat{H}_{\mathrm{XXZ}}$ of Eq.~\eqref{eq:H_XXZ_main} with periodic boundary conditions.
    Left: maximal Bell correlator $\tilde{\Q}$ as a function of the
    exponent $\gamma$ and anisotropy~$\delta$. 
    Right: time traces $\Q(t)$ for the selected pairs $(\gamma,\delta)$ marked (a)--(d) in the phase diagram. Large $\tilde{\Q}$ indicates regimes where the
    SW mapping to OAT with coupling $\chi$ from Eq.~\eqref{eq:H_OAT_XXZ} is accurate.}
  \label{fig:XXZ_phase_diagram}
\end{figure}

We will now switch to the long-range XXZ model from Eq.~\eqref{eq:H_XXZ_main}.
Figure~\ref{fig:XXZ_phase_diagram} summarizes the dependence of Bell
correlations on the microscopic parameters $\gamma$ and $\delta$.
The left panel shows $\tilde{\Q}\equiv\max_t \Q(t)$ scanned over a wide range of points $(\gamma,\delta)$. 
For those marked (a)--(d), the right panel shows the full time dynamics of $\Q(t)$.

In the Heisenberg-like regime $|\delta|\ll1$ and for sufficiently long-range interactions  (small $\gamma$), the gap between the symmetric Dicke manifold
and higher magnon branches is large. Thus, the effective Hamiltonian from Eq.~\eqref{eq:H_OAT_XXZ} accurately captures the dynamics.
In this regime $\tilde{\Q}$ is sizeable over a broad swath of parameter space, and $\Q(t)$ exhibits a single well-defined peak similar to that generated by the  OAT Hamiltonian~\eqref{eq:H_OAT_effective}.
As $\gamma$ increases and interactions become more short-ranged, the effective collective coupling weakens, hence $\tilde{\Q}$ drops. The system crosses from a genuinely
collective regime to a more local one, in which nonlocal correlations are harder to create. For large easy-axis anisotropy $|\delta|\gg1$, the system enters an
Ising-like regime where the effective nonlinearity arises from second-order virtual processes, and sizeable Bell correlations can again be generated in an extended parameter window.
The long-range XXZ chain thus generates strong Bell correlations whenever the magnon gap is large enough to protect the symmetric manifold and the OAT coupling $\chi$ is appreciable.

Another quantum resource of multi-qubit systems that is relevant for ultra-precise metrology is spin-squeezing of the sample~\cite{Wineland1992}. In symmetric systems governed by the collective nonlinearity,
it is typically generated with the OAT dynamics: Starting from the coherent spin state $\ket\uparrow_x^{\otimes N}$,
the nonlinear evolution $\hat H_{\mathrm{eff}}=\chi\hat S_z^2$
shears the quantum uncertainty ellipse in the plane perpendicular to the mean spin,
squeezing the variance along one direction below the standard quantum limit
at the expense of the conjugate direction~\cite{KitagawaUeda1993}.
The degree of squeezing is captured by the spin-squeezing parameter,
\begin{align}\label{eq:xi2_wineland}
  \xi_R^2 = \frac{N\,(\Delta S_\perp^2)_{\mathrm{min}}}{|\langle\hat{\mathbf{S}}\rangle|^2},
\end{align}
where the minimisation runs over all directions perpendicular to the mean
spin $\langle\hat{\mathbf{S}}\rangle$ (see the Appendix).
The spin-squeezing parameter serves as a entanglement witness and a metrological figure of merit accessible 
from collective spin measurements alone~\cite{esteve2008squeezing,Gross2010Nonlinear,Pezze2018RMP}.
Values $\xi_R^2<1$ certify entanglement~\cite{Sorensen2001,Toth2010}
and bound the achievable phase sensitivity in Ramsey interferometry.

\begin{figure}[t!]
  \centering
  \includegraphics[width=\linewidth]{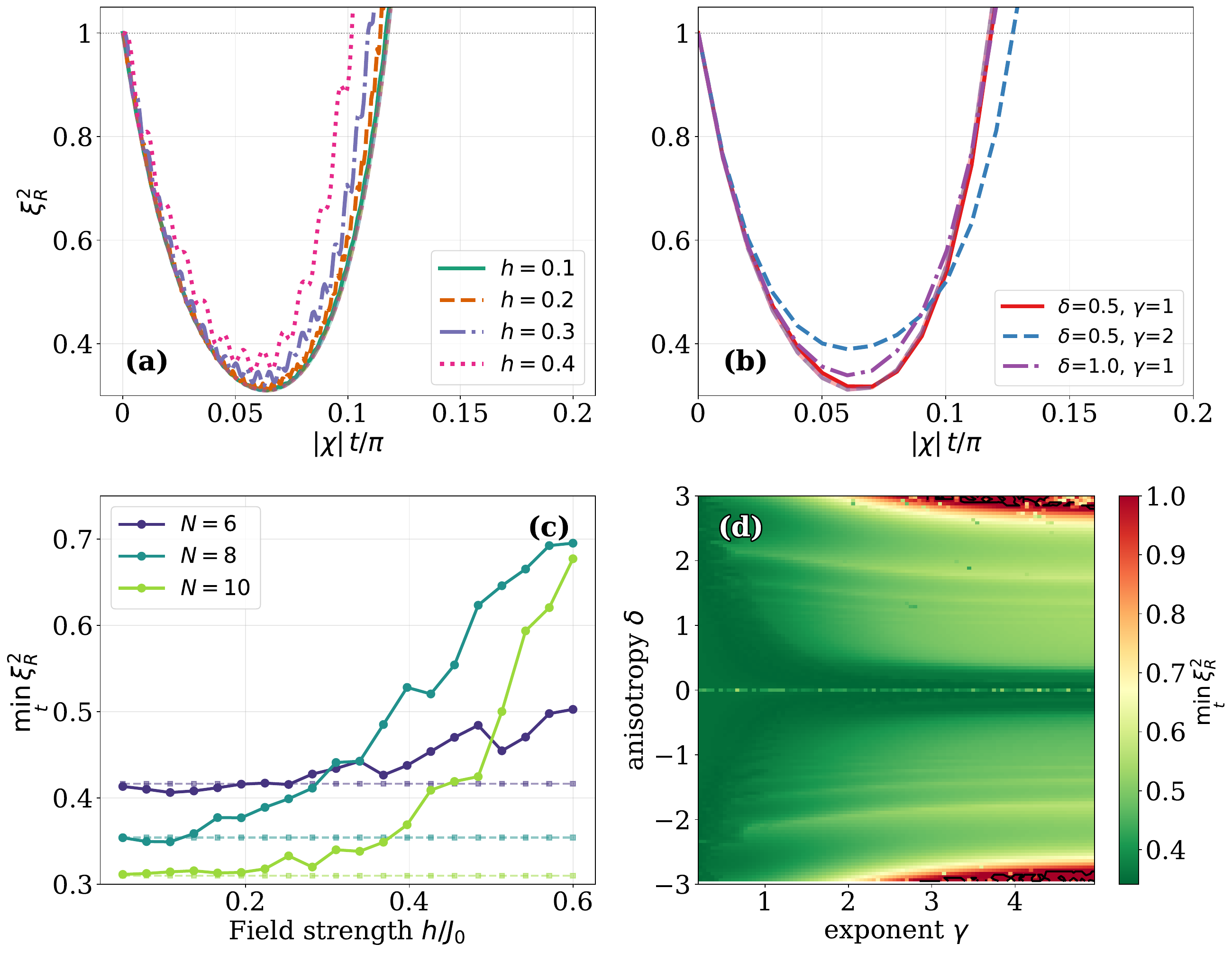}
  \caption{Spin squeezing from microscopic spin chains
    (solid lines) compared with the effective OAT prediction (non-solid lines).
    (a)~Staggered XXX ($N=10$): $\xi_R^2(t)$ for several field strengths $h$.
    (b)~Long-range XXZ ($N=10$): $\xi_R^2(t)$ for representative pairs $(\delta,\gamma)$.
    (c)~Staggered XXX: optimal squeezing $\min_t\xi_R^2$ vs. $h$
    for different $N$ (circles: chain; squares: OAT).
    (d)~Long-range XXZ ($N=8$): $\delta$-$\gamma$ phase diagram of $\min_t\xi_R^2$;
    the black contour marks $\xi_R^2=1$.}
  \label{fig:squeezing}
\end{figure}

Generating spin-squeezed states on digital quantum processors typically requires variational 
optimization of circuit parameters~\cite{Kaubruegger2019VSS,Sun2023Variational1D,Lyu2023VariationalLR,Song2025DipoleSqueezing}. However, this method becomes increasingly costly for larger systems. 
Alternatively, spin squeezing can be achieved through carefully designed fixed gate sequences,~\cite{Carrasco2022ExtremeSS}.
Here, the staggered XXX chain is mapped directly onto a second-order Trotter circuit. This provides a deterministic, nearest-neighbor gate sequence that produces OAT squeezing without variational training.
The same physics can be accessed for the long-range XXZ model through native analog dynamics in analog quantum simulators.
Figure~\ref{fig:squeezing} confirms this picture.
Panels~(a) and~(b) show that $\xi_R^2(t)$ under the full microscopic Hamiltonian closely tracks the OAT prediction (non-solid lines), including the position and depth of the optimal squeezing minimum.
Panel~(c) shows that $\min_t\xi_R^2$ deepens with increasing $h$ in the perturbative regime, consistent with the $\chi\propto h^2$ scaling of Eq.~\eqref{eq:H_OAT_effective}.
Panel~(d) maps the squeezing landscape in the $(\delta,\gamma)$ plane: the region with $\xi_R^2<1$ closely mirrors the Bell-correlation phase diagram of Fig.~\ref{fig:XXZ_phase_diagram}, confirming that both signatures originate from the emergent OAT dynamics.

\section{Discussion and conclusions}\label{sec.disc}
Despite lacking explicit all-to-all couplings, 
we have shown that short-range and power-law interacting spin-$\tfrac12$ chains can generate GHZ coherences and sizeable many-body Bell correlations. 
By applying a Schrieffer-Wolff projection onto the symmetric Dicke manifold, we identified the conditions under which virtual magnon processes across a finite gap produce 
an effective collective twisting of the Lipkin-Meshkov-Glick type.
This mechanism was illustrated in two settings: a staggered nearest-neighbor XXX chain and a long-range XXZ model (see Table in the Appendix for a summary). 
In both cases, the exact dynamics closely follow the effective collective description, producing maximal Bell correlators.
The staggered XXX protocol can be implemented directly on digital quantum hardware. Meanwhile, the long-range XXZ couplings required for the second pathway arise naturally in analog quantum platforms, 
such as superconducting quantum circuits, trapped-ion chains, and Rydberg-dressed atomic ensembles.

We also showed that these correlations can be verified with a single probe qubit that is collectively coupled to the chain. The coherence of the probe encodes the magnetization distribution 
of the rotated many-body state. The Fourier analysis isolates the extremal Dicke coherence that is responsible for the Bell violation. The Fourier weight extracted from the probe 
at the highest mode matches the Bell correlator computed directly.
The same dynamics produce metrologically useful spin squeezing
with $\xi_R^2<1$ over a wide region of parameter space.
Unlike variational quantum circuits~\cite{Kaubruegger2019VSS,Sun2023Variational1D,Lyu2023VariationalLR,Song2025DipoleSqueezing},
the squeezing here requires no classical feedback loop. It follows from the native Hamiltonian evolution with a fixed gate sequence or analog protocol.

\section*{Acknowledgements}
M.P. acknowledges RES resources provided by Barcelona Supercomputing Center in Marenostrum 5 to NNO-2025-3-0004, and MICIU/AEI/10.13039/501100011033/ FEDER, UE.
This work was supported by the National Science Centre, Poland, within the QuantERA II Programme that has received funding from the European Union’s Horizon 2020 
research and innovation programme under Grant Agreement No 101017733, Project No. 2021/03/Y/ST2/00195.
The data and numerical codes that support the findings of this study are available from the corresponding author upon reasonable request.

\onecolumngrid

\appendix

\section{Schrieffer--Wolff projection onto the Dicke manifold}

The SWT~\cite{SchriefferWolff1966} treats systems which exhibit an energy gap $\Delta$ that separates the low-lying part of the spectrum from the high-energy part. 
It allows for a systematic derivation of an effective low-energy ($L$) Hamiltonian, treating the virtual excitations ($E$) as a perturbation. The full Hamiltonian consists of two parts
\begin{align}
  \hat H=\hat H_0+\hat V.
\end{align}
The $\hat H_0$ acts in the two sub-spaces independently, while $\hat V$ provides the coupling. 
The SWT transforms  $\hat H$ into a block-diagonal $\hat H^*$, i.e.,
\begin{align}\label{eq:block}
  \hat H=\left(
  \arraycolsep=1.4pt\def\arraystretch{2.0}
  \begin{array}{c|c}
    \hat H_L & \ \ \lambda\hat V_{LE}\\
    \hline
    \lambda\hat V^\dagger_{LE}\ \  & \hat H_E
  \end{array}
  \right)\ \ \ \xrightarrow{\mathrm{SWT}}\ \ \ \left(
  \arraycolsep=1.4pt\def\arraystretch{1.4}
  \begin{array}{c|c}
    \hat H^*_L & 0\\
    \hline
    0& \hat H^*_E
  \end{array}
  \right)
\end{align}
The second-order correction to the effective low-energy Hamiltonian reads~\cite{endm}
\begin{align}  \label{eq:Heff_eigenbasis}
  \hat H^*_L\equiv\hat\Pi^{(L)}\hat H^*\hat\Pi^{(L)}= -\sum_{nn'm}\frac{\hat\Pi^{(L)}_n\hat V\hat\Pi^{(E)}_m\hat V\hat\Pi^{(L)}_{n'}}{E_n^{(L)}-E_m^{(E)}},
\end{align}
where $\hat\Pi^{(L/E)}_n=\ketbra{\psi_n^{(L/E)}}{\psi_n^{(L/E)}}$ project onto eigenstates of $\hat H_0$ and $E_n^{(L/E)}$ are the corresponding eigen-energies. The derivation of Eq.~\eqref{eq:Heff_eigenbasis} is given in Section~\ref{app:SW_derivation}.

\begin{table*}[t]
\caption{Summary of the Schrieffer--Wolff mapping from microscopic spin chains to OAT dynamics for the two models studied in this work.}\label{tab:recipe}
\begin{ruledtabular}
\begin{tabular}{lll}
 & \textbf{Staggered XXX} & \textbf{Long-range XXZ} \\
\hline
Unperturbed $\hat H_0$ & nearest-neighbour Heisenberg ($J_0$) & Isotropic long-range ($\delta\!=\!0$) or Ising ($|\delta|\!\gg\!1$) \\
Perturbation & staggered field $h_z$ & anisotropy $\delta$ \\
SW order & second & first (projection) \\
Magnon gap $\Delta$ & $2J_0$ & $\tfrac12[(1\!+\!\delta)\tilde J(0)-\tilde J(q_{\min})]$ \\
OAT coupling $\chi$ & $h_z^2/[2J_0(N\!-\!1)]$ & $-\delta\,\tilde J(0)/[2(N\!-\!1)]$ \\
Implementation & digital Trotter circuit & native analog dynamics \\
Platforms & superconducting qubits, neutral-atom arrays & trapped ions, Rydberg tweezers, dipolar systems \\
\end{tabular}
\end{ruledtabular}
\end{table*}

\section{Schrieffer--Wolff effective Hamiltonian for one-body perturbations}\label{app:SW_derivation}

We consider a general isotropic ferromagnetic Hamiltonian $\hat H_0$ perturbed by a one-body field $\hat V=\sum_{j,\alpha}h_j^\alpha\hat S_j^\alpha$.
Because permutation symmetry implies $\hat\Pi\hat S_j^\alpha\hat\Pi=\frac{1}{N}\hat S_\alpha\hat\Pi$, any inhomogeneous field component satisfies $\hat\Pi\hat V_{\mathrm{inh}}\hat\Pi=0$, so the effective collective nonlinearity arises at \emph{second order} in the Schrieffer--Wolff (SW) expansion through virtual magnon excitations across the gap.

The Hilbert space splits into a low-energy subspace $L$ with projector $\hat P$ and its complement $E$ with $\hat Q=\hat{\mathbb I}-\hat P$.
We write $\hat H=\hat H_0+\lambda\hat V$, where $\hat H_0$ is block-diagonal ($\hat P\hat H_0\hat Q=0$), 
and decompose $\hat V=\hat V_{\mathrm d}+\hat V_{\mathrm{od}}$ into block-diagonal and block-off-diagonal parts.
An anti-Hermitian generator $\hat S=\lambda\hat S_1+\lambda^2\hat S_2+\cdots$ is determined such that $\hat H^*=\mathrm{e}^{\hat S}\hat H\,\mathrm{e}^{-\hat S}$ is block-diagonal order by order.
Expanding via Baker--Campbell--Hausdorff,
\begin{align}\label{eq:app_BCH}
  \hat H^*=\hat H+[\hat S,\hat H]+\tfrac{1}{2}[\hat S,[\hat S,\hat H]]+\cdots,
\end{align}
collecting powers of $\lambda$ and imposing the first-order decoupling condition
\begin{align}\label{eq:app_S1_cond}
  \hat V_{\mathrm{od}}+[\hat S_1,\hat H_0]=0
\end{align}
(with $\hat S_1$ purely off-diagonal) removes all block-off-diagonal terms at $O(\lambda)$.
Using $[\hat S_1,[\hat S_1,\hat H_0]]=-[\hat S_1,\hat V_{\mathrm{od}}]$, the 
$O(\lambda^2)$ block-diagonal $LL$-component reduces to
$\tfrac{1}{2}\hat P[\hat S_1,\hat V_{\mathrm{od}}]\hat P$, since both
$\hat P[\hat S_1,\hat V_{\mathrm d}]\hat P$ and $\hat P[\hat S_2,\hat H_0]\hat P$ vanish (the first because a commutator of an off-diagonal with a diagonal operator is off-diagonal, the second by the canonical gauge choice for $\hat S_2$).
Hence
\begin{align}\label{eq:app_Heff_general}
  \hat H_{\mathrm{eff}}=\hat P\hat H_0\hat P+\lambda\hat P\hat V\hat P+\frac{\lambda^2}{2}\hat P[\hat S_1,\hat V_{\mathrm{od}}]\hat P+O(\lambda^3).
\end{align}

Taking matrix elements of Eq.~\eqref{eq:app_S1_cond} between $\ket{n}\in L$ and $\ket{m}\in E$:
\begin{align}\label{eq:app_S1_nm}
  (\hat S_1)_{nm}=\frac{V_{nm}}{E_n^{(L)}-E_m^{(E)}},\qquad
  (\hat S_1)_{mn}=-\frac{V_{mn}^*}{E_n^{(L)}-E_m^{(E)}},
\end{align}
where the second relation follows from $\hat S_1^\dagger=-\hat S_1$.
Evaluating $\bra{n}[\hat S_1,\hat V_{\mathrm{od}}]\ket{n'}$ by inserting a complete set $\sum_{m\in E}\ket{m}\bra{m}$:
\begin{align}\label{eq:app_Heff2_matrix}
  (H_{\mathrm{eff}}^{(2)})_{nn'}=\frac{\lambda^2}{2}\sum_{m\in E}V_{nm}V_{mn'}\times\left(\frac{1}{E_n^{(L)}-E_m^{(E)}}+\frac{1}{E_{n'}^{(L)}-E_m^{(E)}}\right).
\end{align}
When the low-energy manifold is degenerate ($E_n^{(L)}=E_0$ for all $n$), this yields the resolvent form
\begin{align}\label{eq:app_resolvent}
  \hat H_{\mathrm{eff}}=E_0\hat P+\lambda\hat P\hat V\hat P-\lambda^2\hat P\hat V\hat Q\frac{1}{\hat Q\hat H_0\hat Q-E_0}\hat Q\hat V\hat P+O(\lambda^3).
\end{align}

\subsection{Application to the staggered XXX chain}
Consider an isotropic ferromagnetic $\hat H_0$ with a degenerate symmetric Dicke manifold ($S=N/2$, projector $\hat\Pi$) perturbed by a one-body field
$\hat V=\sum_{j,\alpha}h_j^\alpha\hat S_j^\alpha$.
Decomposing $h_j^\alpha=h_0^\alpha+\delta h_j^\alpha$ with $h_0^\alpha=\frac{1}{N}\sum_j h_j^\alpha$ and $\sum_j\delta h_j^\alpha=0$,
the uniform part gives a first-order collective Zeeman term $\hat V_0=\sum_\alpha h_0^\alpha\hat S_\alpha$.
The inhomogeneous part satisfies
\begin{align}\label{eq:app_PiVinhPi}
  \hat\Pi\hat V_{\mathrm{inh}}\hat\Pi=\sum_{j,\alpha}\delta h_j^\alpha\,\hat\Pi\hat S_j^\alpha\hat\Pi
  \stackrel{\eqref{eq:app_Pi_Si}}{=}\frac{1}{N}\sum_\alpha\Bigl(\underbrace{\sum_j\delta h_j^\alpha}_{=\,0}\Bigr)\hat S_\alpha\hat\Pi=0,
\end{align}
so the leading correction from $\hat V_{\mathrm{inh}}$ is second order.

Restricting the resolvent to the one-magnon sector with plane-wave states $\ket{q}$ and energies $\varepsilon(q)$ (the contribution of multi-magnon states is suppressed by additional factors of $h/\Delta$ and enters only at fourth order in the SW expansion; see the convergence discussion below),
we define Fourier modes
$\tilde h_\alpha(q)=\frac{1}{\sqrt{N}}\sum_j\delta h_j^\alpha\,\mathrm{e}^{iqj}$
and
$\hat S_\alpha(q)=\frac{1}{\sqrt{N}}\sum_j\hat S_j^\alpha\,\mathrm{e}^{iqj}$,
so that $\hat V_{\mathrm{inh}}=\sum_\alpha\sum_k\tilde h_\alpha(k)\hat S_\alpha(-k)$. Inserting into the resolvent:
\begin{align}\label{eq:app_Heff2_double_sum}
  \hat H_{\mathrm{eff}}^{(2)}=-\sum_{q\neq 0}\frac{1}{\varepsilon(q)}\sum_{\alpha,\beta}\sum_{k,k'}\tilde h_\alpha(k)\tilde h_\beta(k')\times\hat\Pi\hat S_\alpha(-k)\ket{q}\bra{q}\hat S_\beta(-k')\hat\Pi.
\end{align}
Translational invariance constrains the momenta $(k,k')$ in Eq.~\eqref{eq:app_Heff2_double_sum}.
Under a lattice translation $\hat T_a$, the Fourier spin operators transform as
\begin{align}\label{eq:app_Ta_Sk}
  \hat T_a\hat S_\alpha(k)\hat T_a^\dagger=\mathrm{e}^{ika}\hat S_\alpha(k),
\end{align}
and the momentum eigenstate satisfies $\hat T_a\ket{q}=\mathrm{e}^{iqa}\ket{q}$.
Since $\hat\Pi$ projects onto the fully symmetric (zero-momentum) manifold, it commutes with all translations: $\hat T_a\hat\Pi=\hat\Pi\hat T_a$.

Let $\ket{v}=\hat\Pi\hat S_\alpha(-k)\ket{q}$.
If $\ket{v}\neq 0$, it lies in the range of $\hat\Pi$ and is therefore translation-invariant, $\hat T_a\ket{v}=\ket{v}$.
However, direct evaluation gives
\begin{align}\label{eq:app_v_transform}
  \hat T_a\ket{v}=\hat\Pi(\mathrm{e}^{-ika}\hat S_\alpha(-k))(\mathrm{e}^{iqa}\ket{q})=\mathrm{e}^{i(q-k)a}\ket{v}.
\end{align}
For $\ket{v}\neq 0$, consistency requires $\mathrm{e}^{i(q-k)a}=1$ for all $a$, hence $k=q$.
Analogously, defining $\bra{w}=\bra{q}\hat S_\beta(-k')\hat\Pi$ and applying $\hat T_a^\dagger$ from the right yields the phase $\mathrm{e}^{-i(q+k')a}$, enforcing $k'=-q$.
Only the term $(k,k')=(q,-q)$ survives in Eq.~\eqref{eq:app_Heff2_double_sum}:
\begin{align}\label{eq:app_Heff2_SqSq}
  \hat H_{\mathrm{eff}}^{(2)}=-\sum_{q\neq 0}\frac{1}{\varepsilon(q)}\sum_{\alpha,\beta}\tilde h_\alpha(q)\tilde h_\beta(-q)\,\hat\Pi\hat S_\alpha(-q)\hat S_\beta(q)\hat\Pi.
\end{align}

Expanding the product in real space:
\begin{align}\label{eq:app_SqSq_split}
  \hat S_\alpha(-q)\hat S_\beta(q)=\frac{1}{N}\sum_i\hat S_i^\alpha\hat S_i^\beta+\frac{1}{N}\sum_{i\neq j}\mathrm{e}^{iq(i-j)}\hat S_i^\alpha\hat S_j^\beta.
\end{align}
Project with $\hat\Pi(\cdot)\hat\Pi$.
By permutation symmetry, $\hat\Pi\hat S_i^\alpha\hat S_j^\beta\hat\Pi\equiv\hat A_{\alpha\beta}\hat\Pi$ is independent of the pair $(i,j)$ for $i\neq j$ (see Eq.~\eqref{eq:app_Aab}).
The on-site term is site-independent, contributing $\hat\Pi\hat S_i^\alpha\hat S_i^\beta\hat\Pi$.
For $q\neq 0$, the lattice-sum identity $\sum_{i\neq j}\mathrm{e}^{iq(i-j)}=|\sum_j\mathrm{e}^{iqj}|^2-N=-N$ gives
\begin{align}\label{eq:app_Pi_SqSq_step1}
  \hat\Pi\hat S_\alpha(-q)\hat S_\beta(q)\hat\Pi=\hat\Pi\hat S_i^\alpha\hat S_i^\beta\hat\Pi-\hat A_{\alpha\beta}\hat\Pi.
\end{align}
Inserting the two-body identity~\eqref{eq:app_Aab}:
\begin{align}\label{eq:app_Pi_SqSq_step2}
  \hat\Pi\hat S_\alpha(-q)\hat S_\beta(q)\hat\Pi=\hat\Pi\hat S_i^\alpha\hat S_i^\beta\hat\Pi-\frac{1}{N(N-1)}\bigl(\hat S_\alpha\hat S_\beta-N\hat\Pi\hat S_i^\alpha\hat S_i^\beta\hat\Pi\bigr)\hat\Pi
  =\frac{N}{N-1}\hat\Pi\hat S_i^\alpha\hat S_i^\beta\hat\Pi-\frac{1}{N(N-1)}\hat S_\alpha\hat S_\beta\hat\Pi.
\end{align}
For spin-$1/2$, the algebra $\hat\sigma^\alpha\hat\sigma^\beta=\delta_{\alpha\beta}+i\sum_\gamma\epsilon_{\alpha\beta\gamma}\hat\sigma^\gamma$ gives 
$\hat S_i^\alpha\hat S_i^\beta=\tfrac{1}{4}\delta_{\alpha\beta}+\tfrac{i}{2}\sum_\gamma\epsilon_{\alpha\beta\gamma}\hat S_i^\gamma$.
Projecting with Eq.~\eqref{eq:app_Pi_Si}:
\begin{align}\label{eq:app_onsite_projected}
  \hat\Pi\hat S_i^\alpha\hat S_i^\beta\hat\Pi=\frac{1}{4}\delta_{\alpha\beta}\hat\Pi+\frac{i}{2N}\sum_\gamma\epsilon_{\alpha\beta\gamma}\hat S_\gamma\hat\Pi.
\end{align}
Substituting into the first term of Eq.~\eqref{eq:app_Pi_SqSq_step2}:
\begin{align}\label{eq:app_first_term}
  \frac{N}{N-1}\hat\Pi\hat S_i^\alpha\hat S_i^\beta\hat\Pi=\frac{N}{4(N-1)}\delta_{\alpha\beta}\hat\Pi+\frac{{i}}{2(N-1)}\sum_\gamma\epsilon_{\alpha\beta\gamma}\hat S_\gamma\hat\Pi.
\end{align}
For the second term, decompose $\hat S_\alpha\hat S_\beta=\tfrac{1}{2}\{\hat S_\alpha,\hat S_\beta\}+\tfrac{{i}}{2}\sum_\gamma\epsilon_{\alpha\beta\gamma}\hat S_\gamma$ to separate symmetric and antisymmetric parts.
Combining the linear-in-$\hat S_\gamma$ contributions from both terms:
\begin{align}\label{eq:app_linear_combine}
  \frac{{i}}{2(N-1)}\sum_\gamma\epsilon_{\alpha\beta\gamma}\hat S_\gamma-\frac{{i}}{2N(N-1)}\sum_\gamma\epsilon_{\alpha\beta\gamma}\hat S_\gamma=\frac{{i}}{2N}\sum_\gamma\epsilon_{\alpha\beta\gamma}\hat S_\gamma.
\end{align}
Assembling all contributions yields the identity for $q\neq 0$:
\begin{align}\label{eq:app_Pi_SqSq_final}
  \hat\Pi\hat S_\alpha(-q)\hat S_\beta(q)\hat\Pi&=\frac{N}{4(N-1)}\delta_{\alpha\beta}\hat\Pi-\frac{1}{2N(N-1)}\{\hat S_\alpha,\hat S_\beta\}\hat\Pi+\frac{{i}}{2N}\sum_\gamma\epsilon_{\alpha\beta\gamma}\hat S_\gamma\hat\Pi.
\end{align}

Substituting Eq.~\eqref{eq:app_Pi_SqSq_final} into Eq.~\eqref{eq:app_Heff2_SqSq} and defining
\begin{align}\label{eq:app_Lambda_def}
  \Lambda_{\alpha\beta}=\sum_{q\neq 0}\frac{\tilde h_\alpha(q)\tilde h_\beta(-q)}{\varepsilon(q)},
\end{align}
the second-order effective Hamiltonian separates into three terms of distinct collective structure.

\emph{(i) Quadratic (OAT) term.}
The anticommutator contribution gives
\begin{align}\label{eq:app_Heff_collective}
  \hat H_{\mathrm{eff}}^{(2)}\big|_{\mathrm{quad}}=\sum_{\alpha,\beta}\chi_{\alpha\beta}\,\tfrac{1}{2}\{\hat S_\alpha,\hat S_\beta\}\,\hat\Pi,\qquad
  \chi_{\alpha\beta}=\frac{\Lambda_{\alpha\beta}}{N(N-1)}.
\end{align}
Only the symmetric part of $\chi_{\alpha\beta}$ contributes, since $\{\hat S_\alpha,\hat S_\beta\}$ is symmetric under $\alpha\leftrightarrow\beta$.

\emph{(ii) Linear term.}
The $\hat S_\gamma$ contribution yields
\begin{align}\label{eq:app_Heff2_linear}
  \hat H_{\mathrm{eff}}^{(2)}\big|_{\mathrm{lin}}=\sum_\gamma B_\gamma\hat S_\gamma\hat\Pi,\qquad
  B_\gamma=-\frac{{i}}{2N}\sum_{\alpha,\beta}\epsilon_{\alpha\beta\gamma}\Lambda_{\alpha\beta}.
\end{align}
Because $\epsilon_{\alpha\beta\gamma}$ is antisymmetric, only the antisymmetric part of $\Lambda_{\alpha\beta}$ contributes to $B_\gamma$.
Whenever only a single field component is present (as in the staggered XXX model), $\Lambda_{\alpha\beta}$ is diagonal and the linear term vanishes identically.

\emph{(iii) Constant shift.}
The identity contribution produces
$\hat H_{\mathrm{eff}}^{(2)}\big|_{\mathrm{const}}=-\frac{N}{4(N-1)}(\sum_\alpha\Lambda_{\alpha\alpha})\hat\Pi$,
which renormalizes the ground-state energy but does not affect dynamics within $\hat\Pi$.

For $\hat H_0=-\frac{1}{2}\sum_{i\neq j}J(r_{ij})\hat{\mathbf{S}}_i\cdot\hat{\mathbf{S}}_j$ on a periodic ring,
define $\ket{F}=\ket{\uparrow}^{\otimes N}$ and single-flip states $\ket{j}=\hat S_j^-\ket{F}$.
Using $\hat{\mathbf{S}}_i\cdot\hat{\mathbf{S}}_j=\hat S_i^z\hat S_j^z+\frac{1}{2}(\hat S_i^+\hat S_j^-+\hat S_i^-\hat S_j^+)$
one obtains $\hat H_0\ket{j}=E_0\ket{j}+\frac{1}{2}\sum_{l\neq j}J_{jl}(\ket{j}-\ket{l})$ with $E_0=-\frac{1}{8}\sum_{i\neq j}J(r_{ij})$.
Going to plane waves $\ket{q}=\frac{1}{\sqrt{N}}\sum_j\mathrm{e}^{iqj}\ket{j}$:
\begin{align}\label{eq:app_dispersion}
  \hat H_0\ket{q}=\bigl(E_0+\varepsilon(q)\bigr)\ket{q},\qquad \varepsilon(q)=\tfrac{1}{2}\bigl[\tilde J(0)-\tilde J(q)\bigr],
\end{align}
where $\tilde J(q)=\sum_{r}J(r)\,\mathrm{e}^{iqr}$.
For nearest-neighbor coupling: $\varepsilon(q)=J_0(1-\cos q)$, giving a magnon gap $\Delta=\varepsilon(\pi)=2J_0$.

\subsection{Projection to symmetric manifold}
\label{app:projection_identities}

Let $\hat\Pi$ project onto the fully symmetric Dicke manifold ($S=N/2$).

\paragraph{One-body.}
By permutation symmetry, $\hat\Pi\hat S_i^\alpha\hat\Pi$ is site-independent.
Summing over $i$: $N\hat\Pi\hat S_i^\alpha\hat\Pi=\hat S_\alpha\hat\Pi$, hence
\begin{align}\label{eq:app_Pi_Si}
  \hat\Pi\hat S_i^\alpha\hat\Pi=\frac{1}{N}\hat S_\alpha\hat\Pi.
\end{align}

\paragraph{Two-body ($i\neq j$).}
Similarly, $\hat\Pi\hat S_i^\alpha\hat S_j^\beta\hat\Pi\equiv\hat A_{\alpha\beta}\hat\Pi$ is pair-independent.
From $\hat S_\alpha\hat S_\beta=\sum_i\hat S_i^\alpha\hat S_i^\beta+\sum_{i\neq j}\hat S_i^\alpha\hat S_j^\beta$, projecting and counting $N$ on-site and $N(N-1)$ off-site terms:
\begin{align}\label{eq:app_Aab}
  \hat A_{\alpha\beta}\hat\Pi=\frac{1}{N(N-1)}\bigl(\hat S_\alpha\hat S_\beta-N\hat\Pi\hat S_i^\alpha\hat S_i^\beta\hat\Pi\bigr)\hat\Pi.
\end{align}

\paragraph{On-site spin-$1/2$.}
From $(\hat S_i^z)^2=\frac{1}{4}$, the on-site projected term gives $\hat\Pi\hat S_i^z\hat S_i^z\hat\Pi=\frac{1}{4}\hat\Pi$.
Substituting into Eq.~\eqref{eq:app_Aab} with $\alpha=\beta=z$:
\begin{align}
  \hat A_{zz}\hat\Pi=\frac{1}{N(N-1)}\left(\hat S_z^2-\frac{N}{4}\right)\hat\Pi,
\end{align}
i.e.,
\begin{align}\label{eq:app_Pi_SizSjz}
  \hat\Pi\hat S_i^z\hat S_j^z\hat\Pi=\frac{1}{N(N-1)}\left(\hat S_z^2-\frac{N}{4}\right)\hat\Pi,\qquad i\neq j.
\end{align}
This is the key identity used in the XXZ projection (Section~\ref{app:XXZ_projection}).
An analogous result holds for the $x$ and $y$ components, with $\hat S_z^2$ replaced by $\hat S_x^2$ or $\hat S_y^2$ respectively.

For the staggered model on a ring with even $N$,
$J(r)=J_0$ for $r=1$ and $r=N-1$, giving $\tilde J(q)=2J_0\cos q$ and $\tilde J(0)=2J_0$.
The staggered profile $\delta h_j^z=-h_z(-1)^j$ has Fourier transform
$\tilde h_z(q)=-h_z\sqrt{N}\,\delta_{q,\pi}$
(since $\sum_j(-1)^j\mathrm{e}^{-iqj}=N\delta_{q,\pi}$) and $\tilde h_x=\tilde h_y=0$.
Inserting into Eq.~\eqref{eq:app_Heff_collective}:
\begin{align}\label{eq:app_chi_staggered}
  \chi_{zz}^{(\mathrm{XXX})}=\frac{1}{N(N-1)}\frac{|\tilde h_z(\pi)|^2}{\varepsilon(\pi)}=\frac{1}{N(N-1)}\frac{h_z^2 N}{2J_0}=\frac{h_z^2}{2J_0(N-1)},
\end{align}
reproducing the result from the main text and confirming the general formalism.

The second-order result is controlled by the dimensionless parameter $h/\Delta$.
When the staggered-field amplitude approaches the magnon gap, $h\sim\Delta$, the perturbative hierarchy breaks down: multi-magnon processes become relevant and the effective OAT description ceases to be accurate.
Figure~\ref{fig:app_chi}(b) illustrates this crossover explicitly: for weak fields ($h\ll\Delta$) the numerically extracted $\chi$ follows the analytical $\chi\propto h^2$ scaling, whereas for $h\gtrsim\Delta$ systematic deviations appear.
In practice, the SW expansion remains quantitatively reliable when $h/\Delta\lesssim 0.3$, which is readily satisfied in current experimental platforms where the exchange coupling $J_0$ exceeds the achievable staggered-field amplitudes.

\section{Effective Hamiltonian for the long-range XXZ model}\label{app:XXZ_projection}

In the long-range XXZ model the anisotropy is a \emph{two-body} interaction, $\hat V_\delta\propto\sum_{i\neq j}J(r_{ij})\hat S_i^z\hat S_j^z$.
Unlike the one-body staggered field of Section~\ref{app:SW_derivation}, this perturbation projects nontrivially onto the symmetric manifold, $\hat\Pi\hat V_\delta\hat\Pi\neq 0$, so the effective one-axis-twisting nonlinearity appears already at \emph{first order}.

The full XXZ Hamiltonian reads
\begin{align}\label{eq:app_H_XXZ}
  \hat H_{\mathrm{XXZ}}=-\frac{1}{2}\sum_{i\neq j}J(r_{ij})\left(\hat S_i^x\hat S_j^x+\hat S_i^y\hat S_j^y+(1+\delta)\hat S_i^z\hat S_j^z\right).
\end{align}
We project onto the symmetric manifold, $\hat H_{\mathrm{coll}}=\hat\Pi\hat H_{\mathrm{XXZ}}\hat\Pi$.
Applying the two-body identity $\hat\Pi\hat S_i^\alpha\hat S_j^\alpha\hat\Pi=\frac{1}{N(N-1)}(\hat S_\alpha^2-\frac{N}{4})\hat\Pi$ for $i\neq j$ (Section~\ref{app:projection_identities}) to each component $\alpha=x,y,z$:
\begin{align}\label{eq:app_Hcoll_XXZ}
  \hat H_{\mathrm{coll}}=-\frac{\sum_{i\neq j}J(r_{ij})}{2N(N-1)}\bigl(\hat S_x^2+\hat S_y^2+(1+\delta)\hat S_z^2\bigr)\hat\Pi+\mathrm{const.}
\end{align}
Since $\hat{\mathbf{S}}^2=S(S+1)$ is constant in $S=N/2$, the isotropic part $\hat S_x^2+\hat S_y^2+\hat S_z^2$ produces no dynamics.
Subtracting it, the nontrivial term is
\begin{align}\label{eq:app_chi_OAT_XXZ}
  \hat H_{\mathrm{coll}}\simeq\chi_{zz}^{(\mathrm{XXZ})}\hat S_z^2,\qquad \chi_{zz}^{(\mathrm{XXZ})}=-\frac{\delta}{2}\frac{\tilde J(0)}{N-1},
\end{align}
where $\tilde J(0)=\sum_{r=1}^{N-1}J(r)$ (using $\sum_{i\neq j}J(r_{ij})=N\tilde J(0)$ by translation invariance).
This is an exact operator identity within $\hat\Pi$ and holds for any value of $\delta$.
With Kac normalization (see main text), $\tilde J(0)=J_0$ and the coupling simplifies to $\chi_{zz}^{(\mathrm{XXZ})}=-\delta J_0/[2(N-1)]$.

To obtain the dispersion relation we write $\hat H_{\mathrm{XXZ}}=\hat H_{zz}+\hat H_{xy}$ with $\hat H_{zz}=-\frac{1+\delta}{2}\sum_{i\neq j}J(r_{ij})\hat S_i^z\hat S_j^z$ and $\hat H_{xy}=-\frac{1}{4}\sum_{i\neq j}J(r_{ij})(\hat S_i^+\hat S_j^-+\mathrm{h.c.})$,
the single-flip state $\ket{j}$ sees a diagonal shift from $\hat H_{zz}$: each pair $(j,l)$ contributes $+\frac{1+\delta}{4}J_{jl}$ relative to $\ket{F}$ (because $\hat S_j^z\hat S_l^z$ changes from $+\frac{1}{4}$ to $-\frac{1}{4}$). Summing over both orderings: $\hat H_{zz}\ket{j}=E_F\ket{j}+\frac{1+\delta}{2}\tilde J(0)\ket{j}$.
The flip-flop part gives $\hat H_{xy}\ket{j}=-\frac{1}{2}\sum_{l\neq j}J_{jl}\ket{l}$, identical to the isotropic case.
Combining, $\hat H_{\mathrm{XXZ}}\ket{j}=E_F\ket{j}+\frac{1+\delta}{2}\tilde J(0)\ket{j}-\frac{1}{2}\sum_{l\neq j}J_{jl}\ket{l}$.
Fourier-transforming as before yields
\begin{align}\label{eq:app_XXZ_dispersion}
  \varepsilon_{\mathrm{XXZ}}(q)=\tfrac{1}{2}\bigl[(1+\delta)\tilde J(0)-\tilde J(q)\bigr].
\end{align}
For $\delta>0$ the gap increases with anisotropy, strengthening the protection of the symmetric manifold.

\subsection{Weak anisotropy ($|\delta|\ll 1$): SW with isotropic $\hat H_0$.}
We write $\hat H_{\mathrm{XXZ}}=\hat H_0+\hat V_\delta$ with
$\hat H_0=-\frac{1}{2}\sum_{i\neq j}J(r_{ij})\hat{\mathbf{S}}_i\cdot\hat{\mathbf{S}}_j$ (SU(2)-invariant, exact $S=N/2$ degeneracy) and
$\hat V_\delta=-\frac{\delta}{2}\sum_{i\neq j}J(r_{ij})\hat S_i^z\hat S_j^z$.
Because $\hat V_\delta$ is two-body, $\hat\Pi\hat V_\delta\hat\Pi\neq 0$ (contrast with the one-body case where $\hat\Pi\hat V_{\mathrm{inh}}\hat\Pi=0$).
The leading collective nonlinearity thus appears at first order in the SW expansion:
\begin{align}\label{eq:app_PiVdPi}
  \hat\Pi\hat V_\delta\hat\Pi=-\frac{\delta}{2}\frac{\sum_{i\neq j}J(r_{ij})}{N(N-1)}\left(\hat S_z^2-\frac{N}{4}\right)\hat\Pi,
\end{align}
reproducing Eq.~\eqref{eq:app_chi_OAT_XXZ}.
The isotropic dispersion $\varepsilon(q)=\frac{1}{2}[\tilde J(0)-\tilde J(q)]$ provides the magnon gap protecting the symmetric manifold.
Higher-order SW corrections involve two insertions of $\hat V_\delta$ and generate terms of order $\delta^2$.

\subsection{Strong anisotropy ($|\delta|\gg 1$): SW with Ising $\hat H_0$.}
The natural splitting is
$\hat H_{\mathrm{XXZ}}=\hat H_{0,\mathrm{Ising}}+\hat V_\mathrm{xy}$,
with
\begin{align}
  \hat H_{0,\mathrm{Ising}}=-\frac{1+\delta}{2}\sum_{i\neq j}J(r_{ij})\hat S_i^z\hat S_j^z,\ \ \ \hat V_\mathrm{xy}=-\frac{1}{4}\sum_{i\neq j}J(r_{ij})(\hat S_i^+\hat S_j^-+\text{h.c.}).
\end{align}
The Ising part opens a gap $\Delta\sim(1+\delta)\tilde J(0)$ above the ferromagnetic ground states.
Second-order virtual spin-flip processes across this gap generate an effective OAT nonlinearity consistent with Eq.~\eqref{eq:app_chi_OAT_XXZ}.

The accuracy of the collective description depends on leakage out of $\hat\Pi$ being negligible on the dynamical timescales of interest.
For $\delta=0$, the Hamiltonian is SU(2)-invariant and commutes with $\hat{\mathbf{S}}^2$ for any translation-invariant $J(r_{ij})$; each total-spin sector is then exactly invariant.
For $\delta\neq 0$, SU(2) symmetry is broken and generically $[\hat H_{\mathrm{XXZ}},\hat{\mathbf{S}}^2]\neq 0$, so $\hat\Pi$ is not invariant under the full dynamics.
A notable exception is the permutation-symmetric (all-to-all) case, where $J(r_{ij})=\text{const}$ for all $i\neq j$ and $\hat H_{\mathrm{XXZ}}$ reduces to a polynomial in collective operators, commuting with $\hat{\mathbf{S}}^2$ for any $\delta$.
For finite-range couplings and moderate $\delta$, the finite magnon gap suppresses the leakage rate, ensuring that the projected description remains accurate over dynamic timescales set by $1/\chi$.

\subsection{Comparison between staggered XXX and long-range XXZ models}
The two pathways to an effective OAT nonlinearity identified in this work have distinct perturbative structure.
In the staggered XXX model (Section~\ref{app:SW_derivation}), the nonlinearity arises at \emph{second order} in the one-body perturbation and scales as $\chi^{(\mathrm{field})}\sim h^2/[\Delta(N-1)]$, where $\Delta$ is the magnon gap.
The coupling tensor $\chi_{\alpha\beta}^{(\mathrm{field})}=\Lambda_{\alpha\beta}/[N(N-1)]$ contains explicit energy denominators from virtual magnon excitations.
In the XXZ model, the anisotropy is a two-body interaction that projects nontrivially onto $\hat\Pi$ already at \emph{first order}: $\chi^{(\mathrm{XXZ})}=-\frac{\delta}{2}\tilde J(0)/(N-1)$, with no energy denominators.
The gap is required in both cases, but for different reasons: in the field-induced mechanism it sets the perturbative expansion parameter $h/\Delta$, while in the XXZ case it controls the rate of leakage out of the symmetric manifold.

\section{Numerical verification via exact diagonalization}
\label{app:numerics}

To validate the analytical dispersion relations and effective nonlinearities derived in Sections~\ref{app:SW_derivation} and~\ref{app:XXZ_projection}, we perform exact diagonalization (ED) on finite spin chains.

\begin{figure}[t]
  \centering
  \includegraphics[width=0.6\linewidth]{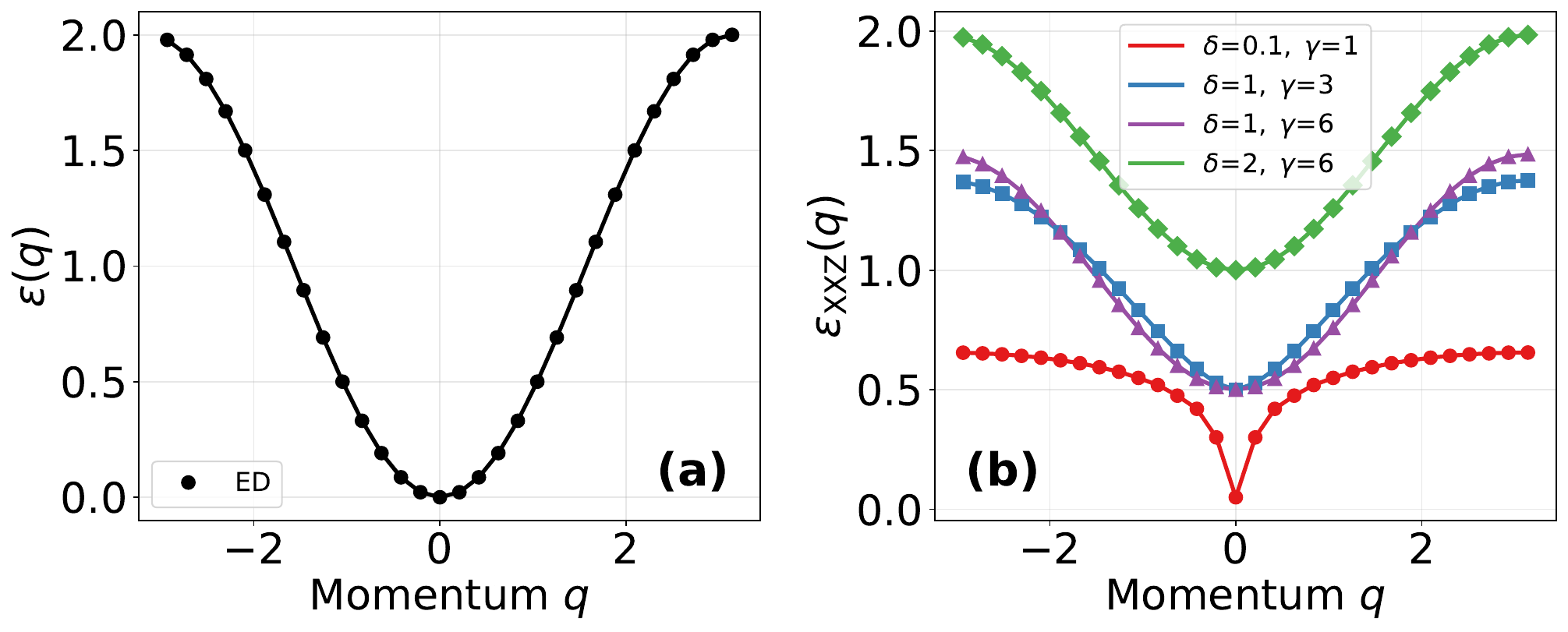}
  \caption{One-magnon dispersion relation: analytical (solid lines) vs.\ exact diagonalization in the single-flip basis (circles), $N=30$.
    (a)~Nearest-neighbour isotropic Heisenberg with $h_z=0$.
    (b)~Long-range XXZ,  for representative $(\delta,\gamma)$ pairs spanning isotropic long-range to strongly anisotropic short-range regimes.}
  \label{fig:app_dispersion}
\end{figure}

On a periodic ring of $N$ spins, the fully polarized state $\ket{F}=\ket{\uparrow}^{\otimes N}$ is the ferromagnetic ground state.
The one-magnon sector is spanned by $N$ single-flip states $\ket{j}=\hat S_j^-\ket{F}$.
Matrix elements of the Hamiltonian in this sector, $H_{jl}=\bra{l}\hat H\ket{j}$, are constructed from the microscopic couplings (diagonal: $(1+\delta)\tilde J(0)/2$; off-diagonal: $-J_{jl}/2$).
Instead of diagonalizing $\hat H$, we project onto the travelling-wave basis
\begin{align}\label{eq:app_travelling_wave}
  \ket{q_k}=\frac{1}{\sqrt{N}}\sum_{j=1}^N\mathrm{e}^{iq_kj}\ket{j},\qquad q_k=\frac{2\pi k}{N},
\end{align}
and compute $\varepsilon_{\mathrm{num}}(q_k)=\bra{q_k}\hat H\ket{q_k}-E_F$.
Figure~\ref{fig:app_dispersion} compares this numerical dispersion with the analytical predictions $\varepsilon(q)=\frac{1}{2}[\tilde J(0)-\tilde J(q)]$ (isotropic) and $\varepsilon_{\mathrm{XXZ}}(q)=\frac{1}{2}[(1+\delta)\tilde J(0)-\tilde J(q)]$, confirming exact agreement.

\begin{figure}[t]
  \centering
  \includegraphics[width=0.6\linewidth]{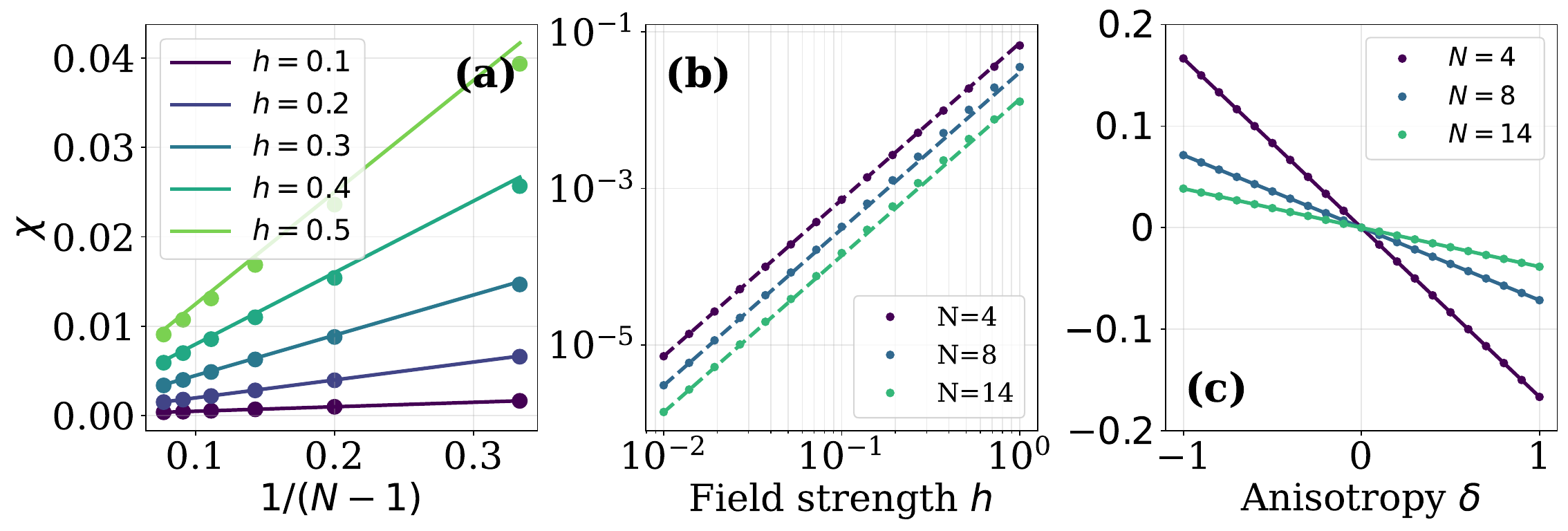}
  \caption{Effective nonlinearity from ED vs.\ analytical predictions.
    (a)~Staggered XXX: $\chi_{zz}^{(\mathrm{XXX})}$ vs.\ $1/(N\!-\!1)$ for several field strengths, confirming the result obtained with the full Hamiltonian to machine precision.
    (b)~ $\chi$ vs.\ field strength (log-log), showing $\chi\propto h^2$ in the perturbative regime and deviations when $h$ approaches the magnon gap $\Delta=2J_0$.
    (c)~Long-range XXZ: $\chi_{zz}^{(\mathrm{XXZ})}$ vs.\ anisotropy $\delta$ at fixed range exponents $\gamma$, confirming the exact linear prediction.
    Lines: theory; markers: ED.}
  \label{fig:app_chi}
\end{figure}

For an effective OAT Hamiltonian $\hat H_{\mathrm{eff}}=\chi\hat S_z^2$, the energy gap between the two highest-$M$ Dicke states satisfies
$\Delta E=E_F-E_W=\chi(N-1)$,
where $E_F=\bra{F}\hat H\ket{F}$ and $E_W$ is the lowest eigenvalue in the single-excitation sector.
Hence $\chi_{\mathrm{num}}=\Delta E/(N-1)$.
Figure~\ref{fig:app_chi} compares this numerical extraction with the analytical predictions for three scenarios.
In panel~(a), the $1/(N-1)$ scaling of $\chi_{zz}^{(\mathrm{XXX})}$ is verified across multiple chain lengths, with the relative deviation $|\chi_{\mathrm{num}}-\chi_{\mathrm{th}}|/\chi_{\mathrm{th}}$ remaining below $10^{-12}$ for all $N$ tested (limited by machine precision), confirming the exact nature of the identity.
Panel~(b) demonstrates the crossover from the perturbative regime ($\chi\propto h^2$) to the non-perturbative regime when the field strength $h$ approaches the magnon gap $\Delta=2J_0$, providing a direct numerical validation of the convergence condition discussed in Section~\ref{app:SW_derivation}.
Panel~(c) shows that the XXZ coupling $\chi_{zz}^{(\mathrm{XXZ})}$ varies linearly with the anisotropy $\delta$ as predicted in the main text, and is independent of $\gamma$ under Kac normalization since $\tilde J(0)=J_0$.

\section{Symmetric-sector fidelity}\label{app:leakage}

The SW transformation maps the microscopic Hamiltonian onto an effective collective model that acts entirely within the symmetric Dicke sector $S=N/2$.
To test how well the dynamics stays confined to this sector, we track the overlap
\begin{align}\label{eq:Fsym}
  F_{\mathrm{sym}}(t) = \langle\psi(t)|\,\hat\Pi\,|\psi(t)\rangle
\end{align}
of the exact time-evolved state $|\psi(t)\rangle = e^{-i\hat{H}t}|{+}x\rangle^{\otimes N}$ with the projector $\hat\Pi$ onto $S=N/2$.
We obtain $\hat\Pi$ by diagonalising $\hat{\mathbf{S}}^2$ and retaining the $(N{+}1)$-dimensional eigenspace with eigenvalue $\tfrac{N}{2}(\tfrac{N}{2}+1)$.

Since $|{+}x\rangle^{\otimes N}$ belongs to the symmetric sector, $F_{\mathrm{sym}}(0)=1$.
Any subsequent decay is driven by the perturbation $\hat V$ (the staggered field $h_z$ or the anisotropy $\delta$), which couples $S=N/2$ to lower-spin subspaces.
Within the SW picture these couplings are suppressed by $\hat V/\Delta$, so $F_{\mathrm{sym}}$ should remain close to unity whenever the perturbative condition holds.

\begin{figure}[t]
  \centering
  \includegraphics[width=0.6\linewidth]{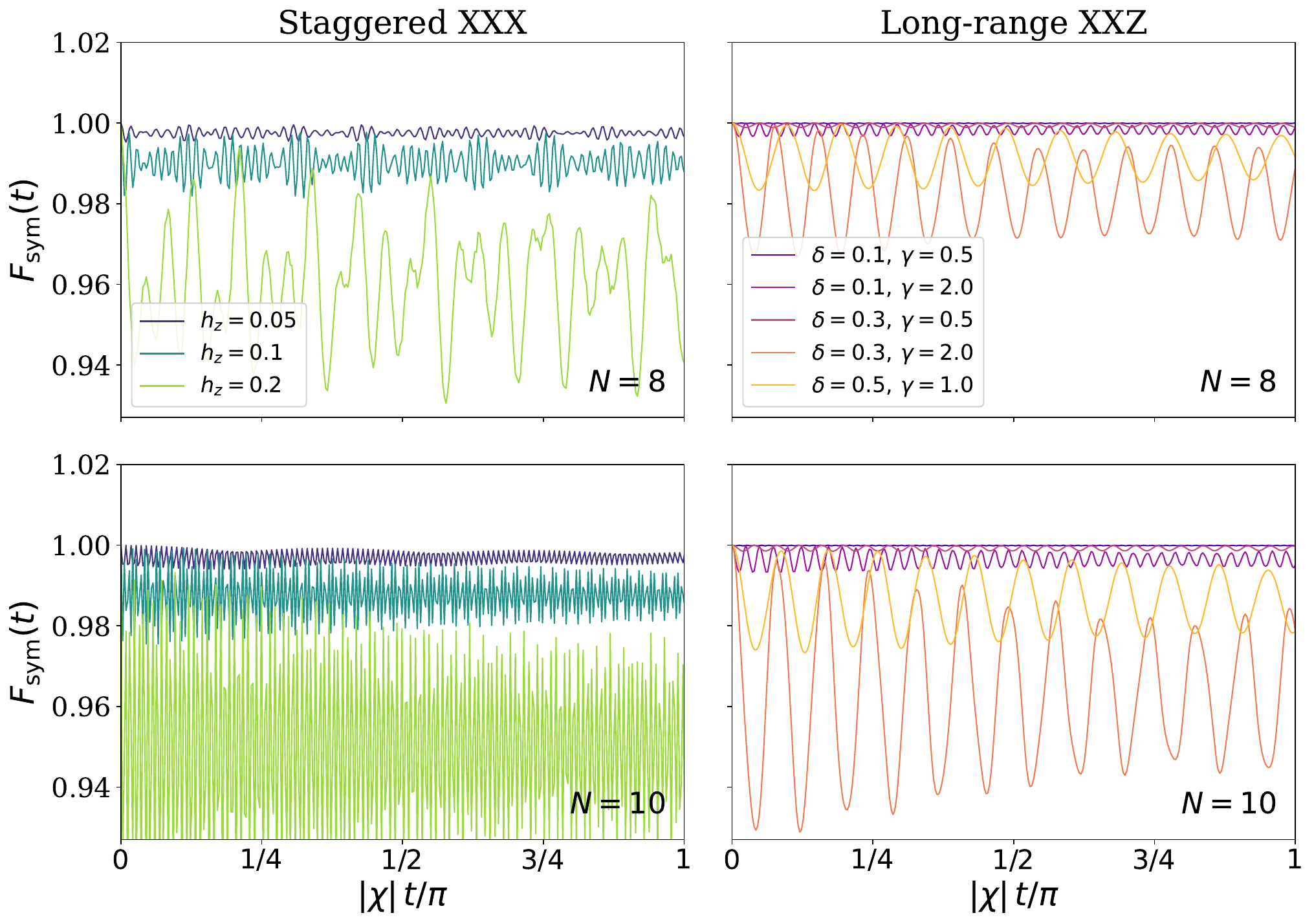}
  \caption{Symmetric-sector fidelity $F_{\mathrm{sym}}(t)$ for the staggered XXX (left) and long-range XXZ (right) models at $N=8$ (top) and $N=10$ (bottom).  Different curves correspond to different perturbation strengths $h_z$ (left) and anisotropy--range pairs $(\delta,\gamma)$ (right).}
  \label{fig:leakage}
\end{figure}

Figure~\ref{fig:leakage} shows $F_{\mathrm{sym}}(t)$ versus the rescaled time $|\chi|\,t/\pi$ for $N=8$ and $10$.
In the staggered XXX model (left column) the fidelity stays above ${\sim}0.95$ for moderate $h_z$ and develops visible oscillations only when $h_z$ approaches the magnon gap $\Delta$.
In the long-range XXZ model (right column), weakly anisotropic long-range couplings ($\delta\lesssim 0.3$, small $\gamma$) keep $F_{\mathrm{sym}}\approx 1$; larger $\delta$ or $\gamma$ shrink the gap and increase leakage.

\section{Analysis tool: Many-body Bell correlator}\label{sec:analysis_tools}

A many-body Bell inequality
is constructed under the assumption of binary outcomes of measurements of two quantities $\sigma^{(k)}_{x/y}=\pm1$, where  $k\in\{1\ldots N\}$ labels 
the particles~\cite{CavalcantiPRL2007,CavalcantiPRA2011,HePRA2011,Niezgoda2020,Niezgoda2021,Chwedeczuk2022,Plodzien2022,Plodzien2024PRA,Plodzien2024PRR,Hamza2024,Plodzien2025ROPP,Plodzien2025PRA,HernandezYanes2025}.
If the $N$-body correlation function $\En\equiv\modsq{\av{\prod_{k=1}^NS_+^{(k)}}}$ is consistent with the postulates of local realism, it satisfies
\begin{align}\label{eq:lhv_2}
  \En=\modsq{\int\!\!d\lambda\,p(\lambda)\prod_{k=1}^NS_+^{(k)}(\lambda)}\leqslant\int\!\!d\lambda\,p(\lambda)\prod_{k=1}^N\modsq{S_+^{(k)}(\lambda)}=2^{-N},
\end{align}
where $p(\lambda)$ is the probability density of a hidden variable, $S_{+}^{(k)}=\frac12(\sigma_x^{(k)}+i\sigma_y^{(k)})$, and
in the last step we used the Cauchy--Schwarz inequality and the fact that $\modsq{S_+^{(k)}(\lambda)}=1/2$ for predetermined binary outcomes.
This is the many-body Bell inequality as its violation defies the postulate of local realism.
For qubits, the classical outcomes are replaced by quantum operators $\hat S_+^{(k)}=\frac12(\hat\sigma_x^{(k)}+i\hat\sigma_y^{(k)})$ (we use $\hat S^\alpha=\hat\sigma^\alpha/2$ throughout), 
and the Bell inequality is
\begin{align} 
  \Enq[\hat\varrho_N]=\modsq{{\rm Tr}\bigg[{\hat{\varrho}_N\,\bigotimes_{k=1}^N\hat S_+^{(k)}}\bigg]}\leqslant2^{-N}.
\end{align}
The above product of $N$ rising operators is non-zero only if it acts on $\ket{\uparrow}^{\otimes N}$, giving $\ket{\downarrow}^{\otimes N}$. Therefore
$\Enq$ is a measure of coherence between these two, which is maximal, $\Enq=\frac14$, for the GHZ state,
\begin{align}
  \ket{\psi}=\frac1{\sqrt2}(\ket{\downarrow}^{\otimes N}+\ket{\uparrow}^{\otimes N}),
\end{align}
and can be expressed as follows
\begin{align}\label{eq:element}
  \Enq[\hat\varrho_N]=\modsq{\varrho_{\frac N2,-\frac N2}}.
\end{align}
Here, $\modsq{\varrho_{\frac N2,-\frac N2}}$ denotes the matrix element of $\hat\varrho_N$ multiplying $\ketbra{\downarrow}{\uparrow}^{\otimes N}$.
It is convenient to introduce $\mathcal Q$ as
\begin{align}\label{eq:def_q}
  \Q(t)=N + \log_2 \Enq[\hat{\varrho}_N(t)],
\end{align}
since now the Bell inequality is equivalent to $\mathcal{Q}\leqslant0$. When it is violated, $\mathcal{Q}>0$, the correlator $\mathcal Q$ certifies the presence of Bell correlations.
Its maximal value, reached with a GHZ state, is $\Q_{\rm max}=N + \log_2\frac14=N-2$.

The OAT protocol is a versatile method for generating states exhibiting Bell correlations of arbitrary depth~\cite{ CavalcantiPRL2007,CavalcantiPRA2011,HePRA2011,
  Niezgoda2020,Niezgoda2021,Chwedeczuk2022,Plodzien2022,Plodzien2024PRA,Plodzien2024PRR,Hamza2024,Plodzien2025ROPP,Plodzien2025PRA,HernandezYanes2024,HernandezYanes2025}.
Hence our interest in mapping the short-range spin systems onto the effective OAT dynamics.

\twocolumngrid

\end{document}